\documentclass[fleqn,10pt]{wlscirep}
\usepackage[utf8]{inputenc}
\usepackage[T1]{fontenc}
\usepackage{subcaption}
\usepackage{siunitx}
\title{A parameter refinement method for Ptychography based on Deep Learning concepts}

\author[1, 2, *]{Francesco Guzzi}
\author[1]{George Kourousias}
\author[1]{Fulvio Billè}
\author[1]{Alessandra Gianoncelli}
\author[1]{Roberto Pugliese}
\author[2]{Sergio Carrato}

\affil[1]{Elettra Sincrotrone Trieste, Italy}
\affil[2]{Image Processing Laboratory (IPL), University of Trieste, Italy}
\affil[*]{francesco.guzzi@elettra.eu}

\begin{abstract}
X-ray ptychography is an advanced computational microscopy technique, which is delivering exceptionally detailed quantitative imaging of biological and nanotechnology specimens, which can be used for high-precision X-ray measurements. However, coarse parametrisation in propagation distance, position errors and partial coherence frequently threaten the experimental viability. In this work, we formally introduce these actors, solving the whole reconstruction as an optimisation problem. A modern deep learning framework was used to autonomously correct the setup incoherences, thus improving the quality of a ptychography reconstruction. Automatic procedures are indeed crucial to reduce the time for a reliable analysis, which has a significant impact on all the fields that use this kind of microscopy. We implemented our algorithm in our software framework, SciComPty, releasing it as open-source. We tested our system on both synthetic datasets, as well as on real data acquired at the TwinMic beamline of the Elettra synchrotron facility.
\end{abstract}
\begin{document}

	\flushbottom
	\maketitle

	\thispagestyle{empty}
\section{Introduction}\label{sect:intro}
In the last decade, computational microscopy methods based on phase retrieval~\cite{Shechtman2015} have been extensively used to investigate the microscopic nature of thin, noncrystalline materials. In~particular, quantitative information is provided by ptychography~\cite{pie2004, Rodenburg2019}, which combines lateral scanning with Coherent Diffraction Imaging (CDI) methods~\cite{Miao1999, robinsonCDI}. Similar to Computed Tomography (CT) \cite{Hounsfield1973, Cormack_1973}, the~technique tries to solve an inverse problem, as~the object is reconstructed from its effects impinged on the incident beam. The~result is a high-detail absorption and quantitative phase image of large specimens~\cite{ptypfe}. In~a transmission setup (e.g., in~a synchrotron beamline microscope~\cite{twinmicstat, giano2021}), this reconstructed object is a complex-valued 2D transmission function $O(x,y)$, which describes the absorption and scattering behaviour~\cite{Dierolf_2010} of the~sample.

\subsection{Iterative Phase~Retrieval}
Typically, an~iterative phase retrieval procedure~\cite{Shechtman2015} is employed to find a solution for $O(x,y)$ \cite{Rodenburg2019}: an {a priori} image formation model is used to simulate the experiment, by~producing synthetic quantities (Figure \ref{fig:simsch}), which depend on the current estimate of all the latent variables, e.g. $O(x,y)$, and the model parameters; it is the comparison (loss function) between simulated and measured quantities that guides the solution, as~the current estimate is iteratively updated (Figure \ref{fig:updsch}) to minimise the error. In~CDI/ptychography, the~quantities of interest are diffraction patterns (Figure \ref{fig:simsch}).

\subsection{The Parameter~Problem}
Ptychography is extremely sensitive to a coarse parameter estimation~\cite{Hue2011, Evo2011}, and~this may result in a severely degraded reconstruction: a long trial and error procedure is typically employed to manually refine these quantities, looking for an output with fewer artefacts. On~the contrary, framing the reconstruction as a gradient-based optimisation process (Figure \ref{fig:updsch}), the~loss function can be written with an explicit dependency on the model parameters, for~which an update function can be calculated. As~the model complexity increases, the~gradient expressions become progressively more difficult to calculate, and~indeed, Automatic Differentiation (AD) methods (also referred to as ``{\em autograd}'') \cite{Bojan2018, Halide2018} applied to ptychography are recently receiving much attention as an effective way to design complex algorithms~\cite{Jiang18, Kandel2019, Du:21}. However, being based on fixed parameters, the~presented reconstruction algorithms can eventually become fragile (except for~\cite{Du:21}), requiring some~interventions.

\subsection{Proposed~Solution}
In the present work, we describe an AD-based ptychography reconstruction algorithm that takes into account many setup parameters within the same optimisation problem. The~procedure was solved {\em entirely} within an AD environment. This was made possible as a loss function was derived by explicitly taking into account {\em all} the setup parameters, which were added to the optimisation pool for a joint regression/reconstruction. It has to be noted that also the probe positions were refined in this way, without~using an expensive Fourier-transform-based approach. Indeed, a~deep-learning-inspired strategy was employed, rooted in the spatial transformer network literature~\cite{STNpaper}. The~software has been released as open-source~\cite{dataautograd}.

\subsection{Manuscript~Organisation}
This paper is organised as follows: in Section~\ref{sect:ptymodel}, we provide a brief introduction to the ptychography forward model, defining the major flaws we wanted to correct, as~well as a brief description of the autograd technology. In~Section~\ref{sect:lossfun}, our computational methodology is described, introducing the designed loss function and the spatial transform components. In~Section~\ref{sect:res}, we present our main results, while Section~\ref{sect:concl} concludes the~paper.

\begin{figure}[htbp]
	\centering
	\includegraphics[width=5 in]{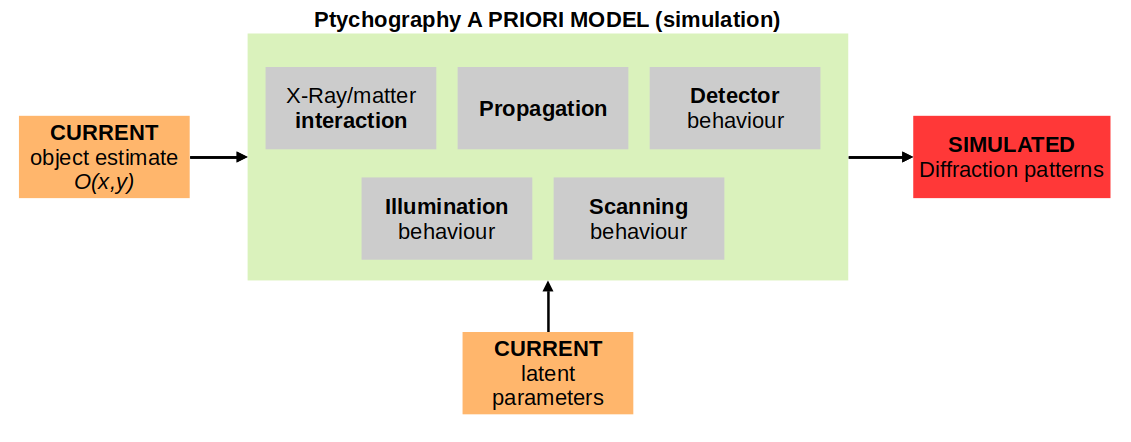}
	\caption{In a computational model (larger box) built on physical insights (embedded boxes), latent variables are used as the input to simulate physical~quantities.}
	\label{fig:simsch}
\end{figure}
\unskip

\begin{figure}[htbp]
	\centering
	\includegraphics[width=5 in]{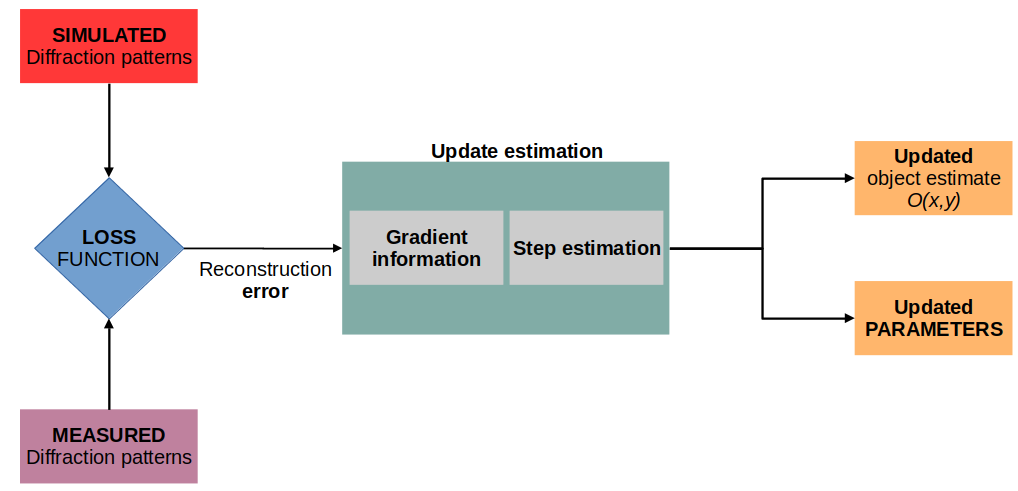}
	\caption{Iterative-gradient-based optimisation: all the latent quantities are updated by using the information from the loss function computed between simulated and measured quantities. The~update estimation is based on the reconstruction error gradient, which is calculated automatically in an automatic differentiation~framework.}
	\label{fig:updsch}
\end{figure}
\unskip

\section{Background} \label{sect:ptymodel}
\begin{figure}[htbp]
	\centering
	\includegraphics[width=5 in]{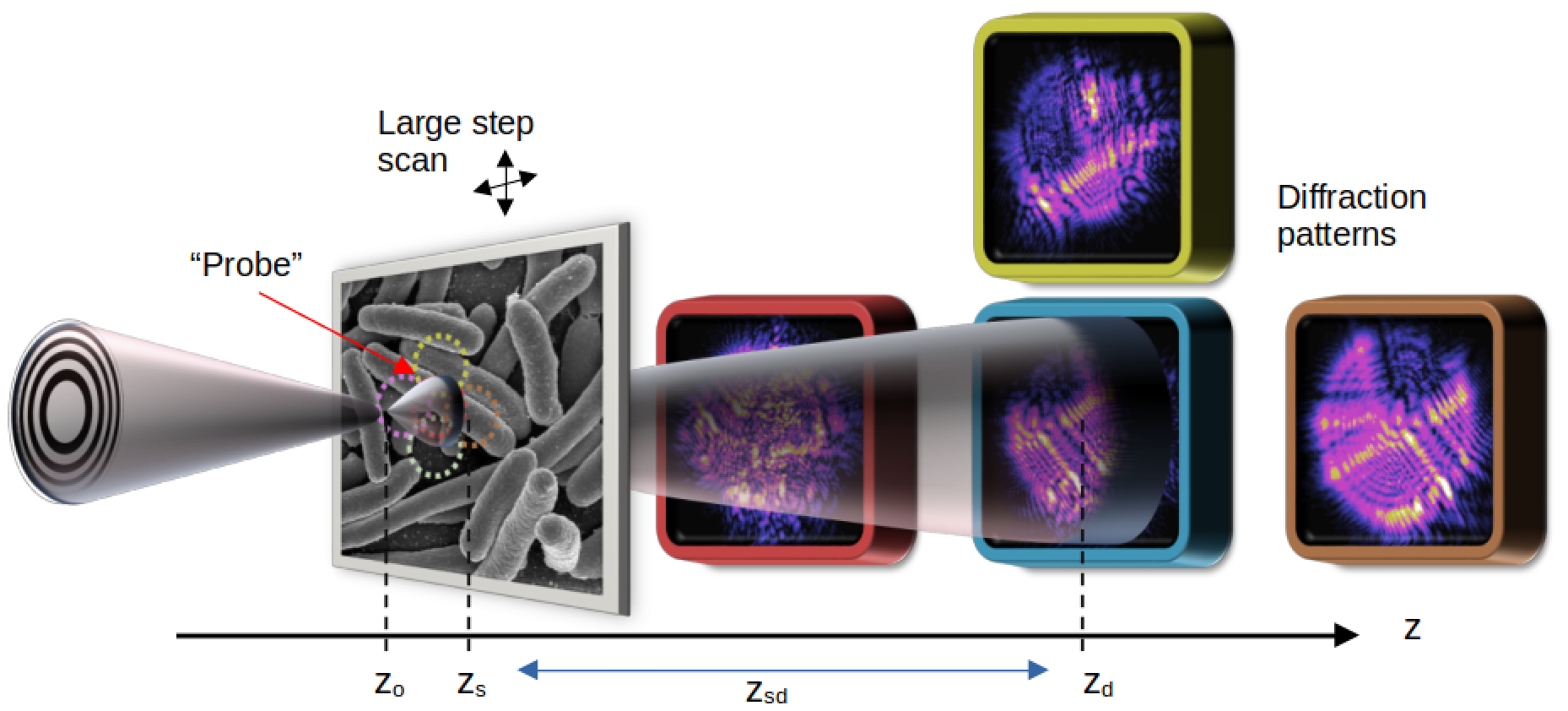}
	\caption{A typical ptychography setup used in synchrotron laboratories: a virtual point source illuminates (probe) a well-defined region on the sample, which is mounted on a motorised stage. The~scattered field intensity is recorded at a distance $z_d$.}
	\label{fig:ptyscheme}
\end{figure}

In ptychography~\cite{Rodenburg2019} (Figure \ref{fig:ptyscheme}), an~extended object is placed onto a sample stage and is illuminated with a conic beam of monochromatic and coherent light. The~radiation illuminates a limited region of the specimen (see the dotted circles in Figure~\ref{fig:ptyscheme}), and~a diffraction pattern is recorded by a detector placed at some distance $z_{sd} = z_d-z_s$ (see Figure~\ref{fig:ptyscheme}). To~reconstruct the object, both the set of $J$ recorded diffraction patterns $I_j$ with $j \in [1...J]$ and the $J$ $(x_j,y_j)$ positions are required. Each $j$th acquisition can be considered independent of the others.
\noindent By illuminating adjacent areas with a high overlap factor, {\em diversity} is introduced to the acquisition, thus creating a robust set of constraints, which greatly improve the convergence of the phase retrieval procedure~\cite{Rodenburg2019}. Diversity is what helps the optimisation problem regress the setup~parameters.

\unskip

\subsection{Ptychography~Model}

The image formation model schematised in Figure~\ref{fig:simsch} can be described more formally by the following expression:
\begin{equation} \label{eq:genforwmodel}
I_j(x, y; z_d) = |D_{z_d-z_s}\{P(x, y; z_s) \cdot O(x_j, y_j; z_s)\}|^2,
\end{equation}
where $P(x,y;z_s)$ is the 2D illumination on the sample plane ($z=z_s$), typically referred to as the ``{\em probe}''; $O(x_j,y_j;z_s)$ is the 2D transmission function of the sample in the local reference system $(x_j,y_j)$, centred at the known $j$th scan position; $D_{z_d-z_s}$ is an operator that describes the observation of the pattern at a known distance $z_d-z_s$.

Knowing the illumination on a region of the object is crucial for the factorisation of the exit wave $\Psi_s(x,y;z_s)$:
\begin{equation} \label{eq:extiwave}
\Psi_s(x,y;z_s) = P(x,y;z_s) \cdot O(x_j,y_j;z_d).
\end{equation}

In a modern ptychography reconstruction, $P(x,y)$ is automatically found thanks to the diversity in the dataset~\cite{thibault2008, Maiden2009, Rodenburg2019}, and~the ``probe retrieval'' procedure has been extended also to the case of a partial coherence~\cite{Whitehead2009, chenJones2012, thibaultmode, batey2014, Rodenburg2019}. Indeed, in~order to take into account the independent propagation of $M$ mutually incoherent probes, Equation~(\ref{eq:genforwmodel}) is modified in the following manner:
\begin{equation} \label{eq:multimode}
I_j(x, y; z_d) = \sum_{p=1}^{M} | D_{z_d-z_s} \{P_p(x, y; z_s) \cdot O(x_j, y_j; z_s) \} | ^2,
\end{equation}

\noindent where the exit wave produced by the modulation of each $P_p(x, z; z_s)$ mode is summed in intensity on the observation plane $z_d$.

\subsection{Parameter~Refinement} \label{sect:parref}
\begin{figure}[htbp]
	\centering
	\includegraphics[width=4.5 in]{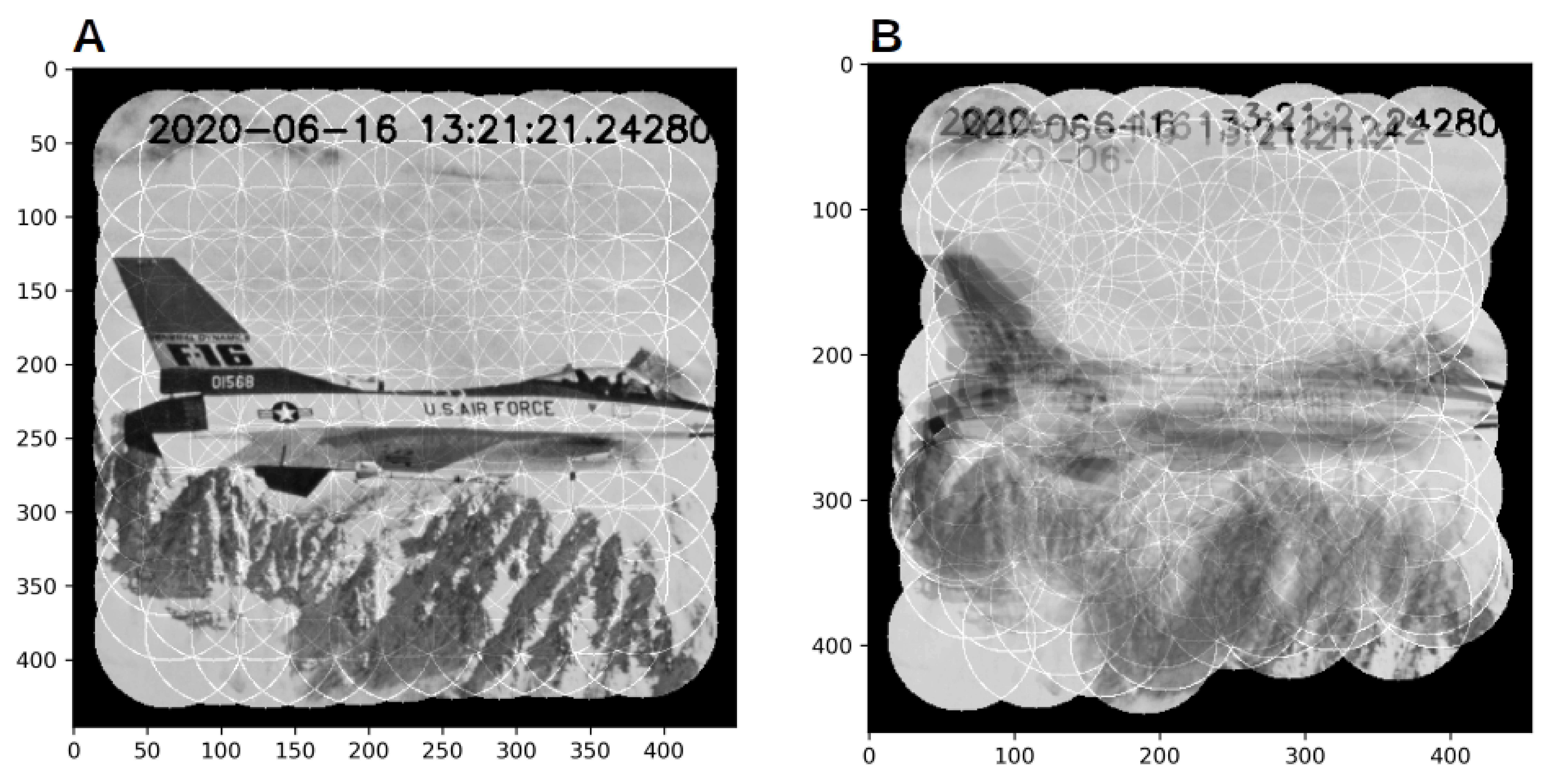}
	\caption{Artificial stitching of the illuminated object ROIs with correct (Panel \textbf{A}) and wrong positions (Panel \textbf{B}): severe artefacts are produced together with a deformation of the total computational box (maximal occupation).}
	\label{fig:poserr}
\end{figure}

We can denote as the ``setup incoherences'' all the deviations of a real setup from the a priori model defined in Equation~(\ref{eq:genforwmodel}). Even if in the literature, many solutions have been proposed, in~most cases, the~problems are typically tackled independently. Partial coherence and mixed-state ptychography have been extensively reviewed (e.g., in~\cite{Li16, Shi2018, Rodenburg2019}); the same is valid for the position refinement problem~\cite{NLpaper,Maiden2012, zhang2013, Tripathy2014, Pynx, guzzi2018,Dwivedi2018}, with~methods that are quite similar to what is performed in, e.g.,~super-resolution imaging~\cite{guarnieri2020} or CT~\cite{guzzi2021}.
\noindent To understand the importance of positions for ptychography, Figure~\ref{fig:poserr} illustrates a slightly exaggerated condition: note that the entire object computational box changes format, and~this is detrimental, especially from the implementation point of~view.

Much more scarce is the literature on axial correction: in~\cite{Evo2011}, an~evolutionary algorithm was used to cope with the uncertainty of the source-to-sample distance in electron ptychography.
In~\cite{lars1}, the~authors proposed using a position refinement scheme also to correct for the axial parameters, as~the latter is responsible for modifying the distances between adjacent probes, while in a work published in January 2021, the~authors of~\cite{lars2} used an autograd environment to directly infer the propagation~distance.

In an even more recent work (March 2021), the~authors in~\cite{Du:21} proposed also a unifying approach to parameter refinement similar to the one used in our manuscript: we genuinely only became aware of these two references while writing the final version of this manuscript. It is undoubted that the automatic differentiation methods (Section \ref{sect:autograd}) are becoming of interest for computational imaging techniques such as ptychography, and~many research teams are approaching the methods as they represent the future of the~technique.

\begin{figure}[htbp]
	\centering
	\includegraphics[width= 4.5 in]{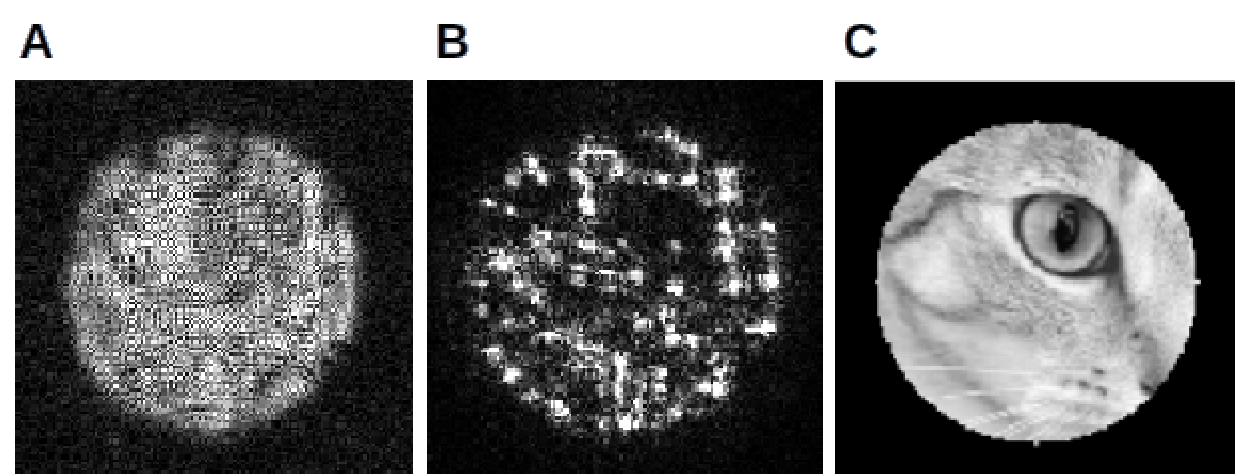}
	\caption{Typical artefacts in a probe retrieval procedure (simulated data) in the case of the wrong propagation distance (Panel \textbf{A}, speckle patterns) and no position refinement (Panel \textbf{B}, cluster of dots). Panel \textbf{C} shows the real illumination (magnitude).}
	\label{fig:illerr}
\end{figure}

Figure~\ref{fig:illerr} shows the effect of the wrong propagation distance on the probe retrieval procedure: in Panel A, speckle-like patterns typically appear; position errors produce instead a typical dotted artefact. An~incorrectly retrieved probe $P(x,y)$ will produce a severely wrong object reconstruction $O(x,y)$.

\unskip

\subsection{Automatic~Differentiation} \label{sect:autograd}
An unconstrained optimisation problem aims at minimising a real-valued {\em loss function} $f(\mathbf{x})$ of $N$ variables. This problem can be formally expressed by the following way:
\begin{equation}
\widetilde{\mathbf{x}} = \operatorname*{argmin}_\mathbf{x} f(\mathbf{x}),
\end{equation}

\noindent where $\widetilde{\mathbf{x}}$ is the sought {\em N}-vector solution. One of the common methods to iteratively minimise the function $f(\mathbf{x})$ is the {\em gradient descent} procedure, which relies on the gradient of the loss function to define an update step:
\begin{equation} \label{eq:graddesc}
\mathbf{x}_{k+1} = \mathbf{x}_k - \alpha \nabla_\mathbf{x} f(\mathbf{x})|_{\mathbf{x}=\mathbf{x}_k}
\end{equation}

\noindent which will provide a new vector estimate $\mathbf{x}_{k+1}$ from the previous estimate $\mathbf{x}_k$, which at convergence will be equal to $\widetilde{\mathbf{x}}$. The~handmade symbolic computation of $\nabla_\mathbf{x} f(\mathbf{x})$
becomes increasingly tedious and error-prone as the complexity of the expression increases. Numerical differentiation automatically provides an estimate of the point derivative of a function by exploiting the central difference scheme, but~while this method is particularly effective for a few dimensions, it becomes progressively slow as $N$ increases. On~the other side, a~Computer Algebra System (CAS) generates a symbolic expression through {\em symbolic computation}, but~often, the~output results in {\em expression swell}. Automatic \mbox{differentiation~\cite{biggs2000,Baydin2018,bart2018}} is a way to provide an accurate gradient calculated at a point, thus lying in between numerical differentiation and handmade calculation.
In one of the currently used methods to automatically compute the gradients~\cite{pytorchpaper}, when a mathematical expression is evaluated, each temporary result constitutes a node in a computational acyclic graph that records the story of the expression, from~the input variables to the generated result. The~gradient is simply calculated following the graph backwards (backward mode differentiation~\cite{bart2018}), from~the results to the input variables, only applying the chain rule to the gradient of each nuclear differentiable operator~\cite{biggs2000}.

\section{Computational~Methodology} \label{sect:lossfun}
The ptychography forward model is defined in terms of a complex probe vector $\mathbf{P} \in \textit{C}^K$ interacting with a complex object transmission function, which can also be arranged as a vector $\mathbf{O} \in \textit{C}^D$.

\subsection{Loss~Function}
Loss functions are typically designed around simple dissimilarity metrics such as quadratic norms, which can be real functions of a complex variable. As~represented in Figure~\ref{fig:updsch}, the~loss function takes into account the simulated ($\widetilde{I_j}$) and real ($I_j$) quantities of the same type (diffraction patterns). When all the conditions defined in Section~\ref{sect:ptymodel} hold, a~loss function $\mathcal{L}$ for all the $j \in [1...J]$ diffraction patterns and positions in the dataset can be written as a {\em data fidelity term}:
\begin{equation}\label{eq:lossfun}
\mathcal{L}(\mathbf{P},\mathbf{O},z) = \sum_{j=1}^{J} \left \| \widetilde{I_j}(\mathbf{P},\mathbf{O},z) - I_j \right \|^{2}.
\end{equation}

As can be seen, Equation~(\ref{eq:lossfun}) is a function of the $2K+2D+1$ real variables, which when stacked up, constitute the optimisation vector pool. The~simulated diffraction pattern $\widetilde{I_j}$ relative to the current $j$th computational box is calculated by Equation~(\ref{eq:genforwmodel}) or by its extension to a multimode illumination (Equation (\ref{eq:multimode})). The~square root of the recorded data can be computed once for all the diffractions to fasten the implementation. $D_z$ is the angular spectrum propagator~\cite{paganin2006} defined by the expression:
\begin{equation} \label{eq:angspect}
\Psi_d (x,y; z_d) = \mathcal{F}^{-1}\{ \mathcal{F}\{ \Psi_s(x,y;z_s)\} \cdot \mathcal{F}\{h(x,y,z)\}\},
\end{equation}

\noindent which relates the input field $\Psi_s(x,y; z_s)$ (defined in Equation~(\ref{eq:extiwave})) to the output field at the detector plane $\Psi_d(x,y; z_d)$. Fixing the wavelength $\lambda$ of the incident radiation, the~2D Fourier transform of the propagation filter $h$ is defined by~\cite{paganin2006}:
\begin{equation}
\mathcal{F}\{h(x,y,z)\} = H(f_x,f_y, z) = e^{jkz \cdot\sqrt{1-(\lambda f_x)^2 - (\lambda f_x)^2}}.
\end{equation}

\subsection{Complex-Valued~AD}
Current DL autograd tools are not conceived of to work with complex numbers, so a basic complex library needs to be written; the natural way to introduce them is to just add an extra dimension to each 2D tensor and incorporate the real and the imaginary part in the same object, basically duplicating the number of actual variables. The~automatic backward operation is completely acceptable for this kind of custom-made data type, and~the resulting gradient is simply:
\begin{equation} \label{eq:wirtinger}
\operatorname{grad} f(\mathbf{x}) = 2\nabla_x^* f(\mathbf{x}) = \frac{\partial f}{\partial \mathbf{x}^*} = ( \frac{\partial f}{\partial \mathbf{a}} +j \frac{\partial f}{\partial \mathbf{b}} )
\end{equation}

\noindent where $\mathbf{x} = \mathbf{a}+j\mathbf{b}$. The~actual gradient is represented in the same complex data type of the variables, made of a real and an imaginary part. As~can be seen in Equation~(\ref{eq:wirtinger}), the~result of the automatic differentiation can be written in the Wirtinger formalism~\cite{wirtingerbook}, just as the derivative with respect to the conjugate of the differentiation variable (except for the constant), which is the typical gradient expression exploited for functions of complex~variables.

\subsection{Regularisation}
To increase the quality of the reconstruction, the~data fidelity term in a loss function is usually paired with a regularisation term (the method of Lagrange multipliers), whose role is to penalise {ad~hoc} solutions in the parameter space: a pixel value ought to fit the physical nature of the model beneath, without~only accommodating the dissimilarity measure; this translates into a mere energy-conservation constraint. Other regularisation methods can be employed, especially the one based on priors on the image frequencies; however, they can be more difficult to tune and require a different forward model. Different from other works (e.g., \cite{Kandel2019, Du:21}), in~this work, especially for $\mathbf{O}$ and $\mathbf{P}$, we decided to use energy-based regularisation, paired only with a large penalisation of quantities out of the desired range. In this way, we practically constrained the minimisation even if a nonconstrained optimisation framework was~used.

\subsection{Spatial Transform~Layer} \label{sect:stn}
If as seen in Equations~(\ref{eq:lossfun}) and (\ref{eq:angspect}), writing a mathematical expression in the parameters $\mathbf{O}$, $\mathbf{P}$ and $z$ directly permits the gradient generation, the same does not hold for the spatial shift correction defined on a discrete sampling grid. Indeed, in~the typical ptychography reconstruction algorithm, the~computational box for a given $j$th position is defined by exploiting a simple crop operator on a 2D tensor. While this method is simple and computationally fast (pointer arithmetics), it is not differentiable, as~integer differentiation is nonsense. In~this work, an~approach borrowed from the DL community was then~explored.

Convolutional neural networks are not invariant to geometric transforms applied to their input. To~cope with this problem, a~spatial transform layer~\cite{STNpaper} is introduced; the new learnable model is trained by inferring the spatial transformation that, applied to the input feature map, maximises the task metric. This approach greatly improves classification and recognition performances (e.g., in~face recognition~\cite{guzzi2020}).

\begin{figure}[htbp]
	\centering
	\includegraphics[width=5 in]{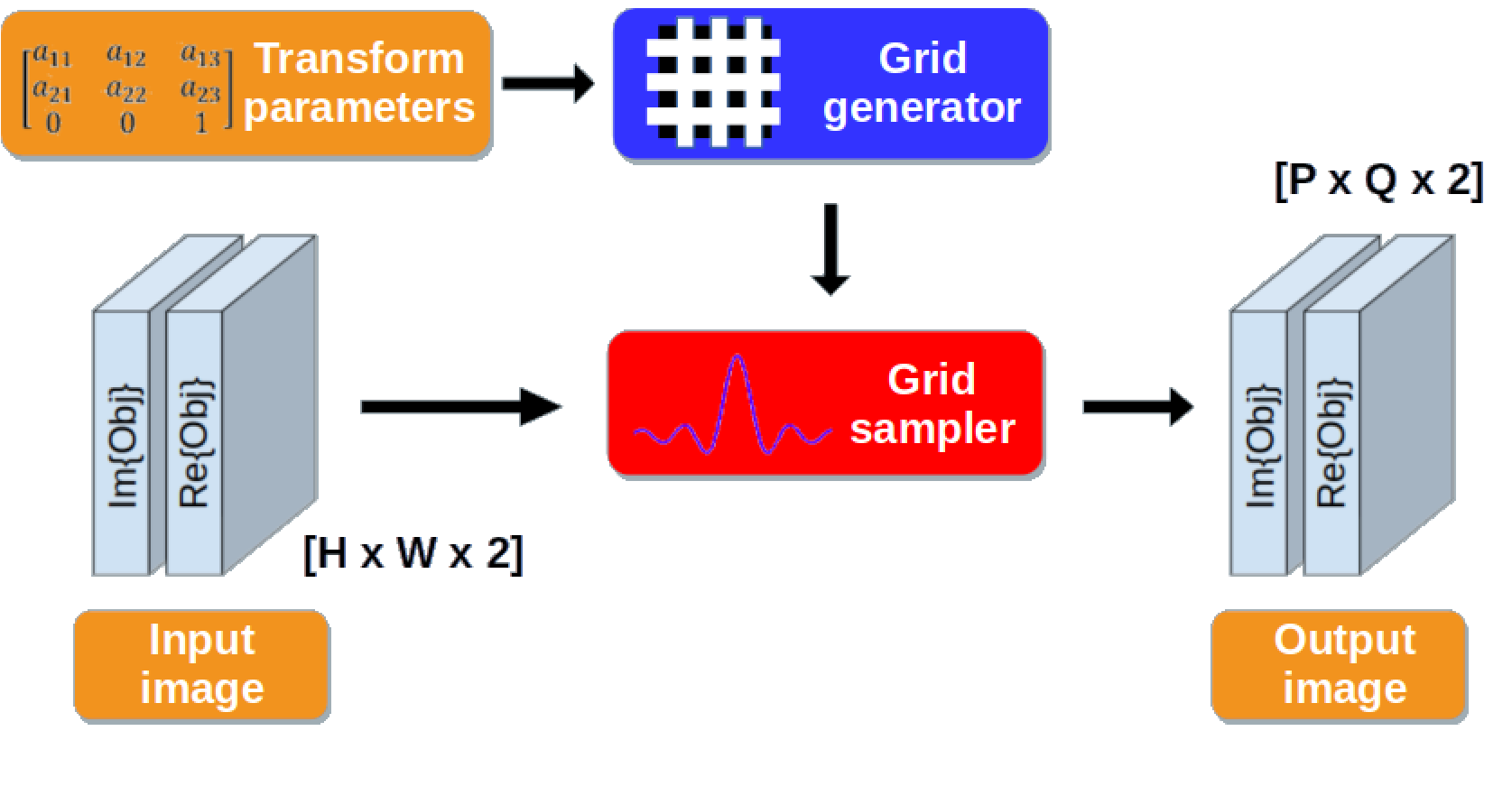}
	\caption{Schematics of the differentiable components used in our method. For~each positions vector, a~grid generator takes as the input the corresponding shift transform expressed in an affine transform formalism; the sampling grid is then generated using this information. The~object is then sampled at each coordinate defined by the sampling grid, producing the correct cropped region of the~object.}
	\label{fig:stnscheme}
\end{figure}

In this work, instead, the~parameters of the affine transform were directly learned as the position refinement coefficients, which minimise the objective function (\mbox{Equation (\ref{eq:lossfun})}) during the reconstruction. To~do so, two components were used (Figure \ref{fig:stnscheme}): (I) a grid generator and (II) a grid sampler. The~first element transforms a regular input sampling grid $(x_i,y_i)^T$ into an output coordinates grid $(x_o,y_o)^T$, by~applying an affine transform to the former; this mapping (Equation  \ref{eq:affine}) is fully defined by six degrees of freedom, but~as concerns rigid translation, only the last column $(t_x, t_y)^T$ is optimised. The~scaling factors $(s_x, s_y)^T$ are thus constants used to crop the central region.
\begin{equation} \label{eq:affine}
\begin{pmatrix} x_o \\ y_o \end{pmatrix}_j = \begin{pmatrix} s_x & 0 & t_x \\ 0 & s_y & t_y \end{pmatrix}_j \begin{pmatrix} x_i \\ y_i \end{pmatrix}_j
\end{equation}

Then, for~each $j$th diffraction pattern and each $j$th shift-vector $(t_x,t_y)^T$, the~$j$th output grid $(x_o,y_o)^T$ is generated. From~the latter, the~grid sampler thus outputs a warped version of the object. Following the formalism introduced in~\cite{STNpaper}, the~cropped portion of the object $V_j$ is generated from the entire object $U$ by:
\begin{equation}
V_j = \sum _{h=1}^{H} \sum _{w=1}^{W} U_{w,h} \cdot K(w-x_{j}, h-y_{j})
\end{equation}

\textls[-15]{A well-performed sampling (with an antialiasing filter) adds a regularisation term that can help during the optimisation. The~bilinear sampling exploits a triangular (separable) kernel $K$, defined in Equation~(\ref{eq:bili}), which does not present dangerous overshooting~artefacts:}
\begin{equation} \label{eq:bili}
K(w-x,h-y) = K_1(w-x) \cdot K_1(h-y)
\end{equation}

\noindent where:
\begin{equation}
K_1(t) = \operatorname*{max}(0, 1-|t|).
\end{equation}

Figure~\ref{fig:stnscheme} shows the structure of the proposed approach, applied to the ptychography framework. The~same spatial transform is enforced on the two channels of the tensor that represent the real and imaginary parts of the cropped region of the object. Within~this setup, gradient propagation is possible because derivative expressions can be calculated with respect to both the affine grid and the output pixel values~\cite{STNpaper}.

\unskip

\section{Results and~Discussion} \label{sect:res}

In this section, reconstructions obtained from a soft X-ray experiment are presented. Additional details on the performance analysis for many synthetic datasets can be found instead in the Supplementary Material of this publication. The~reconstruction obtained through the proposed method is confronted with the output of the EPIE~\cite{Maiden2009} and RPIE~\cite{rpie} algorithms. The~virtual propagation distance of $0.037$ mm was chosen by reconstructing at many $z$ and selecting the best reconstruction. All the computational experiments were written on PyTorch 1.2~\cite{pytorchpaper} and executed on a computer equipped with an Intel Xeon (R) E3-1245 v5 CPU running at 3.50 GHz. The~entire code was implemented on a GPU (Nvidia Quadro P2000), which is essential for this heavy-duty computational~imaging.

The imaging experiment was performed at the TwinMic spectromicroscopy beamline~\cite{twinmicstat, giano2021} at the Elettra synchrotron facility. TwinMic can operate in three imaging modalities: (I) STXM; (II) full-field TXM/CDI; (III) scanning CDI (ptychography). Clearly, the~latter (the one used for this work) is obtained by combining the optic setup of the second modality and the control of the sample stage from the first~one.

Similar to other ptychography experiments performed at the beamline (e.g., \cite{jones2013, giano2021}), X-ray data were collected using a 1020 eV X-ray synchrotron beam~\cite{twinmicstat} focused with a \SI{600}{\micro\metre} diameter Fresnel Zone Plate (FZP) with an outer zone width of \SI{50}{\nano\metre}. The~zone plate was placed approximately at 2 m downstream a \SI{25}{\micro\metre} aperture that defines a secondary source. This is crucial to increase the beam coherence, at~the expenses of brightness. A~Peltier-cooled Charge-Coupled Device (CCD) detector (Princeton MT-MTE) with 1300 $\times$ 1340 \SI{20x20}{\micro\metre} px$^2$ was placed roughly 72 cm downstream of the FZP. According to the Abbe theory~\cite{jones2013}, the~limit of the resolution for coherent illumination is $\Gamma = 0.82 \lambda / NA$ = 50 nm. The~resolution in Fresnel CDI is a function of the experimental geometry~\cite{Quiney2010}, namely the distance from the focal point of the FZP to the detector and the physical size of the detector itself, rather than the focusing~optics.

From the ptychography configuration point of view, the~situation is similar to the one presented in~\cite{Stockmar2013}, where a point source (obtained by idealising the focus of an FZP through an order sorting aperture) illuminates the sample. Following the approach of~\cite{Stockmar2013}, during~the data analysis, the~beam was parallelised by using the Fresnel scaling theorem~\cite{paganin2006}. In~the following experiment, as~pointed out in Section~\ref{sect:ptymodel}, the~sample was considered sufficiently thin~\cite{thibault2008, epie3d} to be modelled by a multiplicative complex transmission function $O$ defined by the expression:
\begin{equation}
O(\mathbf{r}) = e^{j\frac{2\pi}{\lambda}[n(\mathbf r)-1]t(\mathbf r))} = e^{j\frac{2\pi}{\lambda}t(\mathbf r)[-\delta(\mathbf r)+j\beta(\mathbf{r})]}
\end{equation}

\noindent where $\mathbf{r}$ is the planar coordinate on the sample plane, $n(\mathbf{r})$ is the complex refraction index and $t(\mathbf{r})$ a real $R^2 \mapsto R^1$ function defining the local thickness. From~the reconstruction-inferred $O(\mathbf{r})$ (the objective of the reconstruction), the~magnitude map corresponds to:
\begin{equation}
\ln|O(\mathbf r)| = \ln\sqrt{I(\mathbf r)} = \ln \sqrt{I_o(\mathbf r)}-\frac{2\pi}{\lambda}\beta(\mathbf r)\cdot t(\mathbf r)
\end{equation}

\noindent where $I_o$ is the flat field intensity and $I$ the sample intensity. The~phase map instead corresponds to:
\begin{equation}
\mathrm{arg}\{O(\mathbf r)\} = \phi(\mathbf r)=-\frac {2\pi}{\lambda}\delta(\mathbf r) \cdot t(\mathbf r).
\end{equation}

If the properties of the material ($\beta$ and $\delta$) are known, it is simple to infer its thickness and vice~versa.

Diffraction data were acquired in the form of a 16-bit multipage tiff file. The~nominal positions were directly acquired from the shift vectors provided to the control system of the mechanical stage. A~series of dark field images was acquired for the dark field~correction.

\begin{figure}[htbp]


\centering
\begin{subfigure}[H]{\textwidth}
	\centering
	\includegraphics[width=\linewidth]{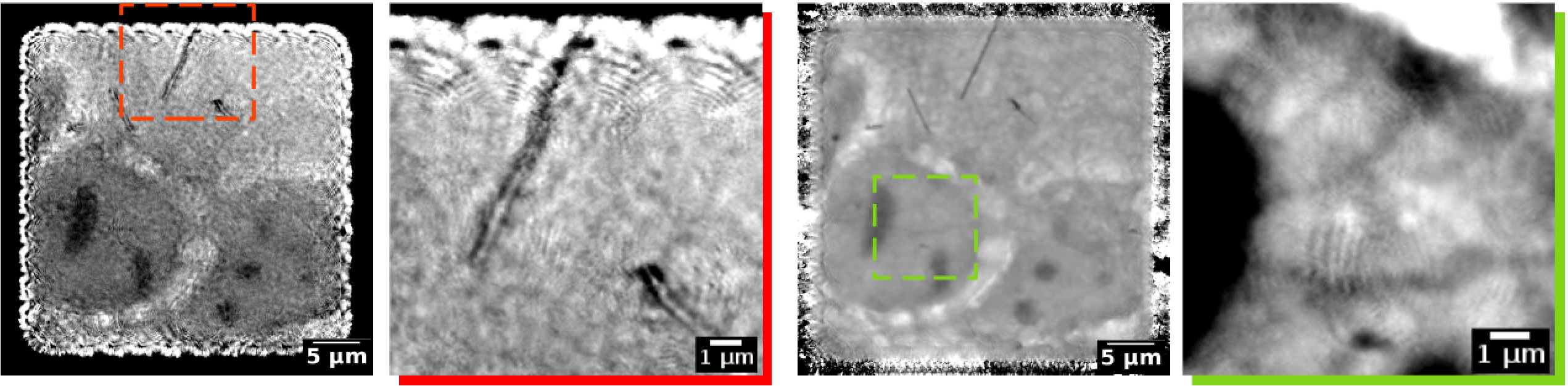}
	\caption{}
	\label{fig:ePIErecon}
\end{subfigure}

\vspace{2em}

\centering
\begin{subfigure}[H]{\textwidth}
	\centering
	\includegraphics[width=\linewidth]{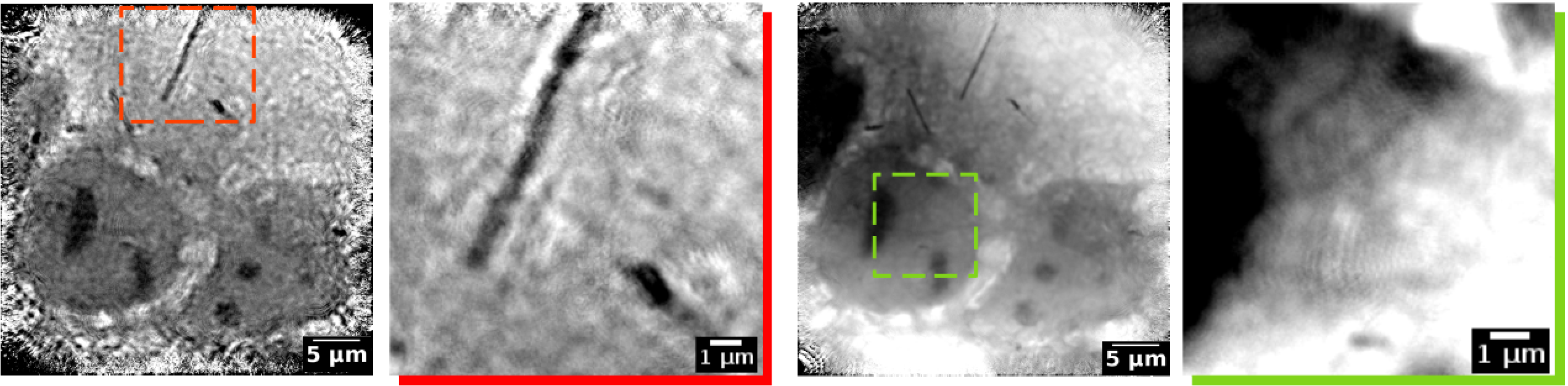}
	\caption{}
	\label{fig:RPIErecon}
\end{subfigure}

\vspace{2em}

\centering
\begin{subfigure}[H]{\textwidth}
	\centering
	\includegraphics[width=\linewidth]{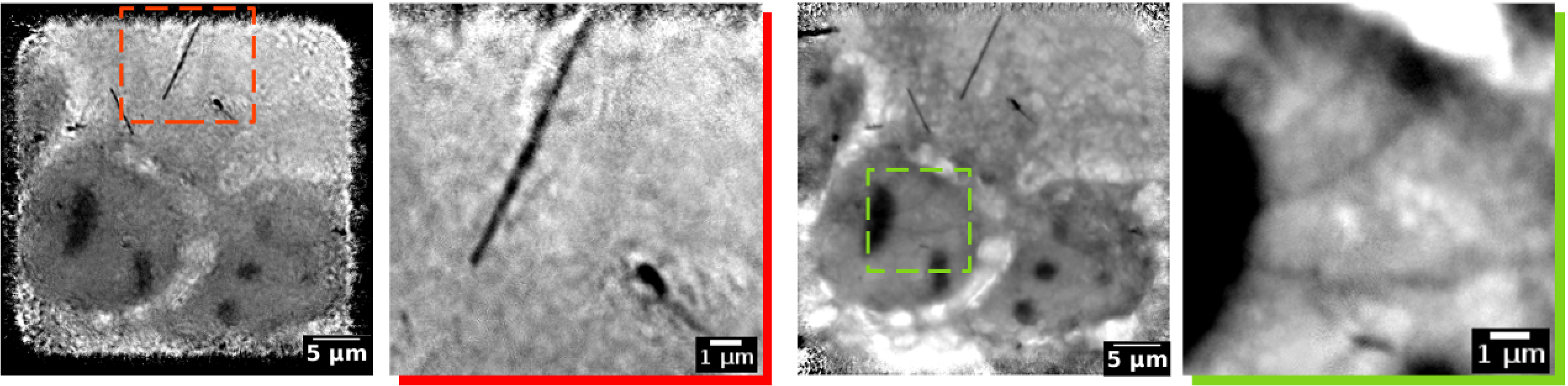}
	\caption{}
	\label{fig:scicom_recon}
\end{subfigure}

\vspace{2em}

\centering
\begin{subfigure}[htbp]{\textwidth}
	\centering
	\includegraphics[width=0.4 \textwidth]{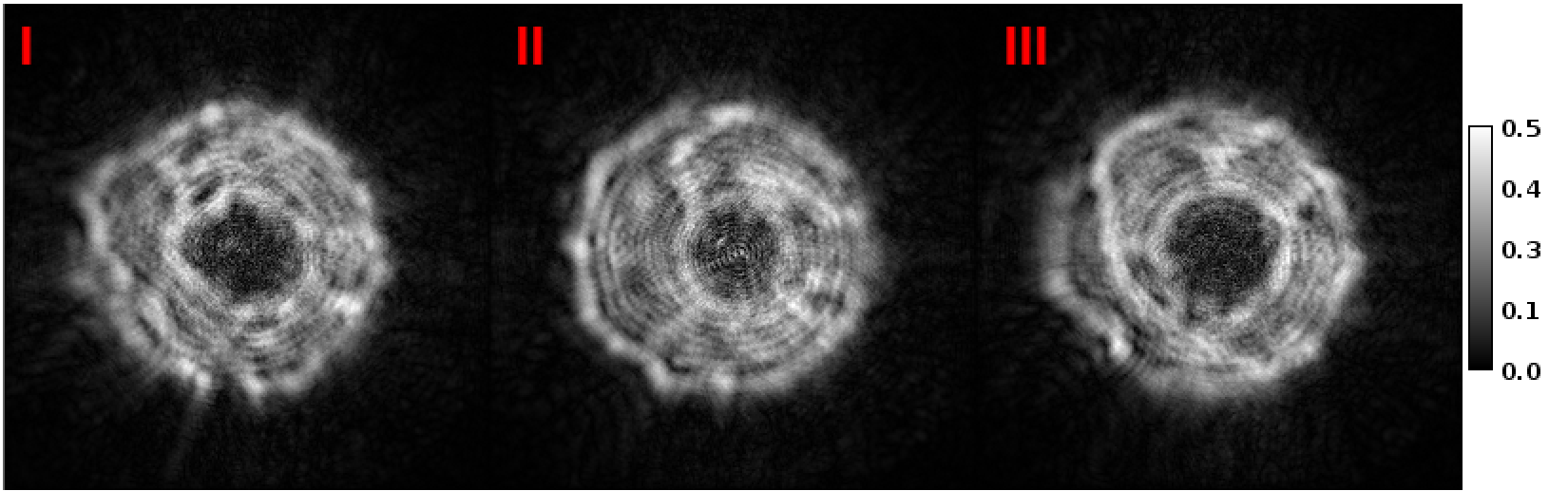}
	\caption{}
	\label{fig:scicom_reconill}
\end{subfigure}

\vspace{1em}

\caption{Ptychography reconstructions of MET cells exposed to asbestos: the proposed algorithm (Panel d) provides the sharpest reconstruction, as~can be seen from the insets. Panel d shows the retrieved multimode~illumination. \mbox{(\textbf{a}) Multimode}~EPIE. (\textbf{b}) Single-mode~RPIE. (\textbf{c}) SciComPty autograd (proposed method). (\textbf{d}) SciComPty \mbox{autograd~illumination.}}
\label{fig:softxray}
\end{figure}

Figure~\ref{fig:softxray} shows a group of chemically fixed mesothelial cells: Mesenchymal--Epithelial Transition (Met5A) cells were grown on silicon nitride windows and exposed to asbestos fibres~\cite{giano2021}. The~absorbing diagonal bar is indeed an asbestos fibre included in the sample. The~reconstructed sample is shown in magnitude (second columns) and phase (third column), where the first and the fourth ones are the magnification of the red (magnitude) and green (phase) area denoted in the figure. Each row shows the reconstruction obtained with a different algorithm (EPIE, RPIE and the proposed method ``SciComPty autodiff''). Due to the fact that for each algorithm, many iterations (10,000) were needed for this dataset, in~order to reduce the computation time, each diffraction pattern was scaled to 256 $\times$ 256 px, giving a resulting pixel size of roughly 36 nm (\textasciitilde 4 $\times$ 9 nm = 36 nm) on a \mbox{846 $\times$ 847 px} reconstructed image. As~in all the simulated experiments, for~each diffraction pattern, the~correct value of the padding was inferred by the propagation routine, taking into account the wavelength and the current propagation distance~value.

\begin{figure}[htbp]
	\centering
	\includegraphics[width= 3.2 in]{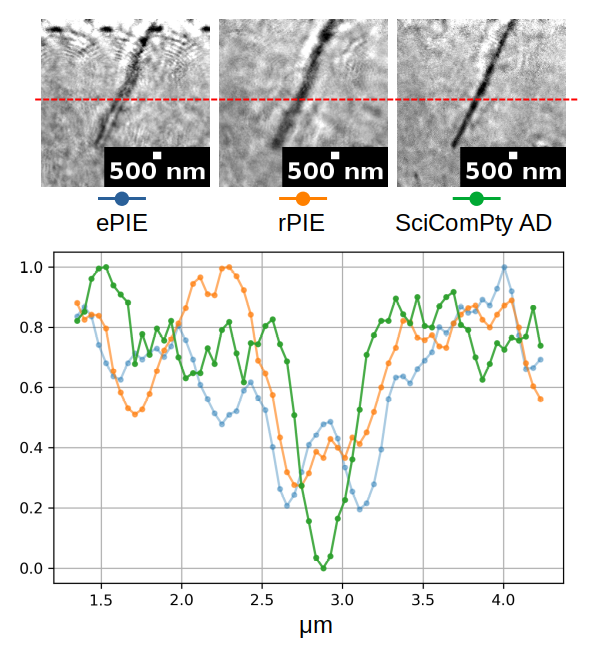}
	\caption{Line profile for each of the reconstructions in Figure~\ref{fig:softxray}.}
	\label{fig:lineprof}
\end{figure}

Observing the quality of the results, the~proposed method (Fig. \ref{fig:scicom_recon}) clearly surpassed all the aforementioned ones: Figure~\ref{fig:softxray} shows the red insets of the top left fibre, which was correctly reconstructed with the highest resolution only by the proposed algorithm; in a multimode EPIE reconstruction, many ringing artefacts are visible. RPIE~\cite{rpie} provided the best result among the typical reconstruction algorithms, with~an object with a large field of view, which extended also into sparse sampled areas. The~proposed method (fourth row) reconstructed the fibre and the cell at the highest resolution in both magnitude and phase (see Figure~\ref{fig:lineprof}). A~second inset (green colour) shows the texture in the phase reconstruction, where again, the~proposed method outperformed the others; cell structures were corrupted by fewer artefacts and visible in their entire~length.

\nointerlineskip


\begin{figure}[htbp]
	\includegraphics[width=0.95\linewidth]{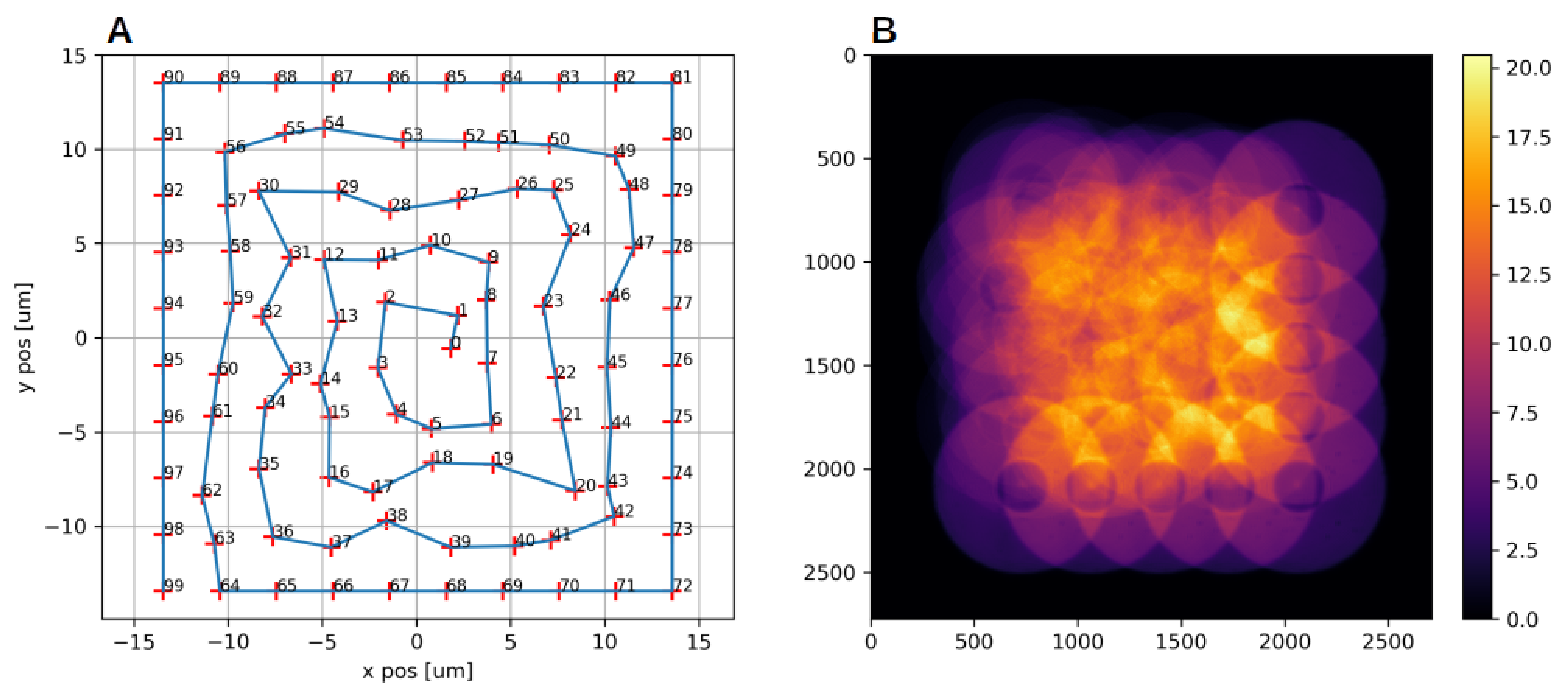}
	\caption{Panel \textbf{A} shows the final scan positions in micrometres. The~corresponding sample density mask is represented in Panel \textbf{B}, which denotes a high sampling density, especially in the centre (higher overlap).}
	\label{fig:scanpatternreal}
\end{figure}

The higher reconstruction quality can be directly attributed to the combined action of both the automatic inference of the virtual propagation distance (Fresnel scaling theorem) of 0.24 mm instead of the 0.37 mm (as obtained by an exhaustive manual search for the other algorithms) and the use of an advanced optimisation algorithm (Adam~\cite{Adam}) in which the choice of the batch size represents a new hyperparameter that can be tuned in grain steps (therefore easily). In~all the other methods, the~position refinement was also enabled. The~final scan positions retrieved by the algorithm are shown in Figure~\ref{fig:scanpatternreal}. We stress that the use of a DL-inspired position refinement routine was here not in competition with other methods, but~it was necessary to carry out the reconstruction within an AD~framework.

The downscaling is required not only for speed reasons, but~also due to the high GPU memory consumption, which is currently a drawback of the method: as the gradient are calculated per batch, increasing the batch size produces a faster computation (less gradients are calculated for the entire set of diffraction patterns), but~the memory consumption is greatly~increased.

The proposed method thus provides a good reconstruction of both the magnitude and phase of the object transmission function. Figure~\ref{fig:lineprof} shows the line profile for each magnitude reconstruction with an FWHM resolution that is clearly twice better than the other~methods.

\section{Conclusions} \label{sect:concl}

Automatic Differentiation (AD) methods are rapidly growing in popularity in the scientific computing world. This is especially true for difficult computational imaging problems such as ptychography or CT. In~this paper, an~AD-optimisation-based ptychography reconstruction algorithm was presented, which retrieves at the same time the object, the~illumination and the set of setup-related quantities. We proposed a solution to solve at the same time for partial coherence, position errors and sensitivity to setup incoherences. In~this way, accurate X-ray measurements become possible, even in the presence of large setup incoherences. This kind of automatic refinement not only allows improving the viability of a ptychography experiment, but~also is particularly crucial to reduce the time for a reliable analysis, which currently is highly hand-tuned.
Extended tests were performed on synthetic datasets and on a real soft-X-ray dataset acquired at the Elettra TwinMic spectromicroscopy beamline, resulting in a very noticeable quality increase. We implemented our algorithm in our modular ptychography software framework, SciComPty, which is provided to the research community as open-source~\cite{dataautograd}.

\section*{Appendix: experiments on synthetic data} \label{sect:synth}

\noindent A ptychography simulator, implemented in the SciComPty modular framework, is used to create 5 synthetic datasets, starting from a set of standard test images and setup parameters which spans different propagation distances, from 0.065 m to 0.2 m. The wavelength is kept constant at 1 nm. All the computational experiments are written on PyTorch 1.2 \cite{pytorchpaper} and executed on a computer equipped with an Intel Xeon (R) E3-1245 v5 CPU running at 3.50 GHz. The entire code is implemented on GPU (Nvidia Quadro P2000), which is essential for this heavy-duty computational imaging. During the tests on synthetic data, the resolution for each diffraction pattern is limited to 128x128 pixels to reduce the computation time. The overlap factor is kept constant at around 70\%, producing a scan pattern of 11x11 positions. A known random jitter is added to the ideal grid-based scan pattern to avoid the "raster scan pathology" \cite{thibault2008}. The resulting total object size has a field of view of around 512x512 pixels (roughly 200 $\mu$ m). Within the context of this paper, the term "epoch", which is used interchangeably with "iteration", defines how many times the entire set of $J$ diffraction patterns is processed through the algorithm.

\noindent The use of an autograd environment allows to easily experiment with the batch-size parameters: EPIE and DM are completely antipodal in this sense, as the first is a sequential algorithm (stochastic gradient descent), while the second employs all the measured data at once (gradient descent). The batch size hyper-parameter (here set at 5 probes) allows to span between the two worlds. Within this framework, new first-order optimisers such as Adam \cite{Adam} become readily available, providing a considerable acceleration to the plain old gradient descent method described by eq. 4 and 5 in the main text. 

\noindent To investigate the effects of our correction routine, for each dataset the propagation distance is initialised to a value corrupted by a 30\% error, while each position is perturbed by a random jitter with a standard deviation of 10 pixels. Both the propagation distance and the position vectors are added to the optimisation pool. Apart from eye inspection, to validate the method, reconstruction quality can be analysed observing the behaviour of (I) the optimisation dissimilarity (eq. $6$ in the main text), calculated between the simulated and the measured diffraction pattern, and (II) a truth-aware similarity metric, SSIM \cite{ssim}, which obviously can be applied for simulated data only. In the latter case, the positions error and the propagation distance estimate can be monitored at each iteration, producing informative graphics which are the base for the following analysis.

\begin{figure}[hbtp]
	\centering
	\includegraphics[width=4 in]{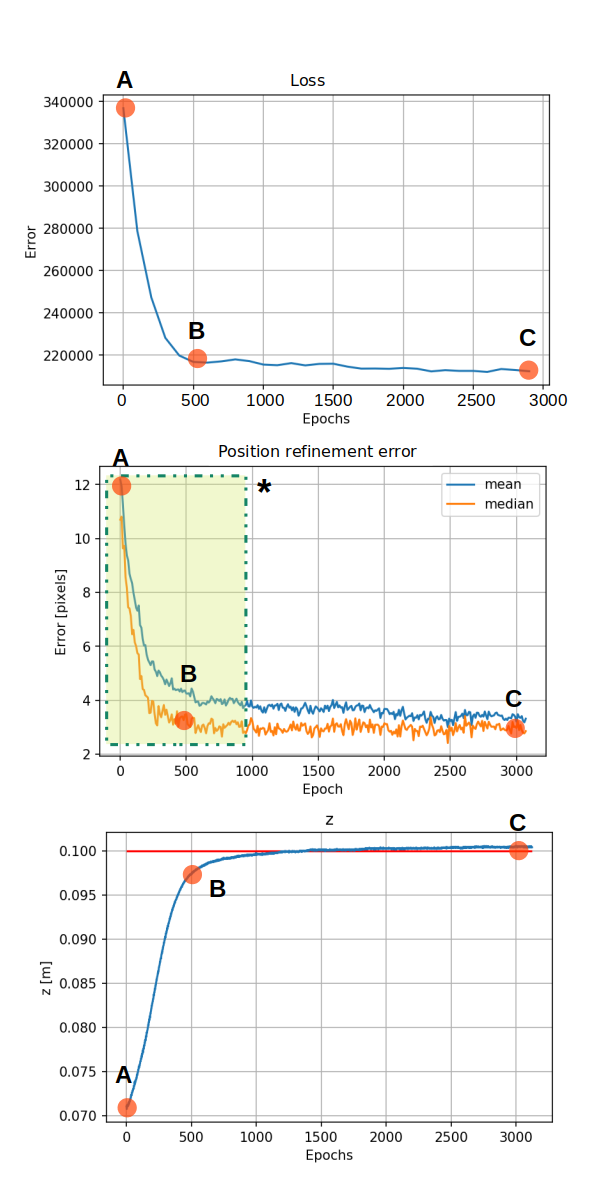}
	\caption{The convergence of our combined optimisation is analysed by observing: the loss value, the ground truth position refinement error and the propagation distance $z$ inferred by the algorithm. In the bottom panel, the correct distance is denoted by the red horizontal line at $z= 0.1$m}
	\label{fig:combinederror}
\end{figure}

\begin{figure}[hbtp]
	\centering
	\includegraphics[scale=1.7]{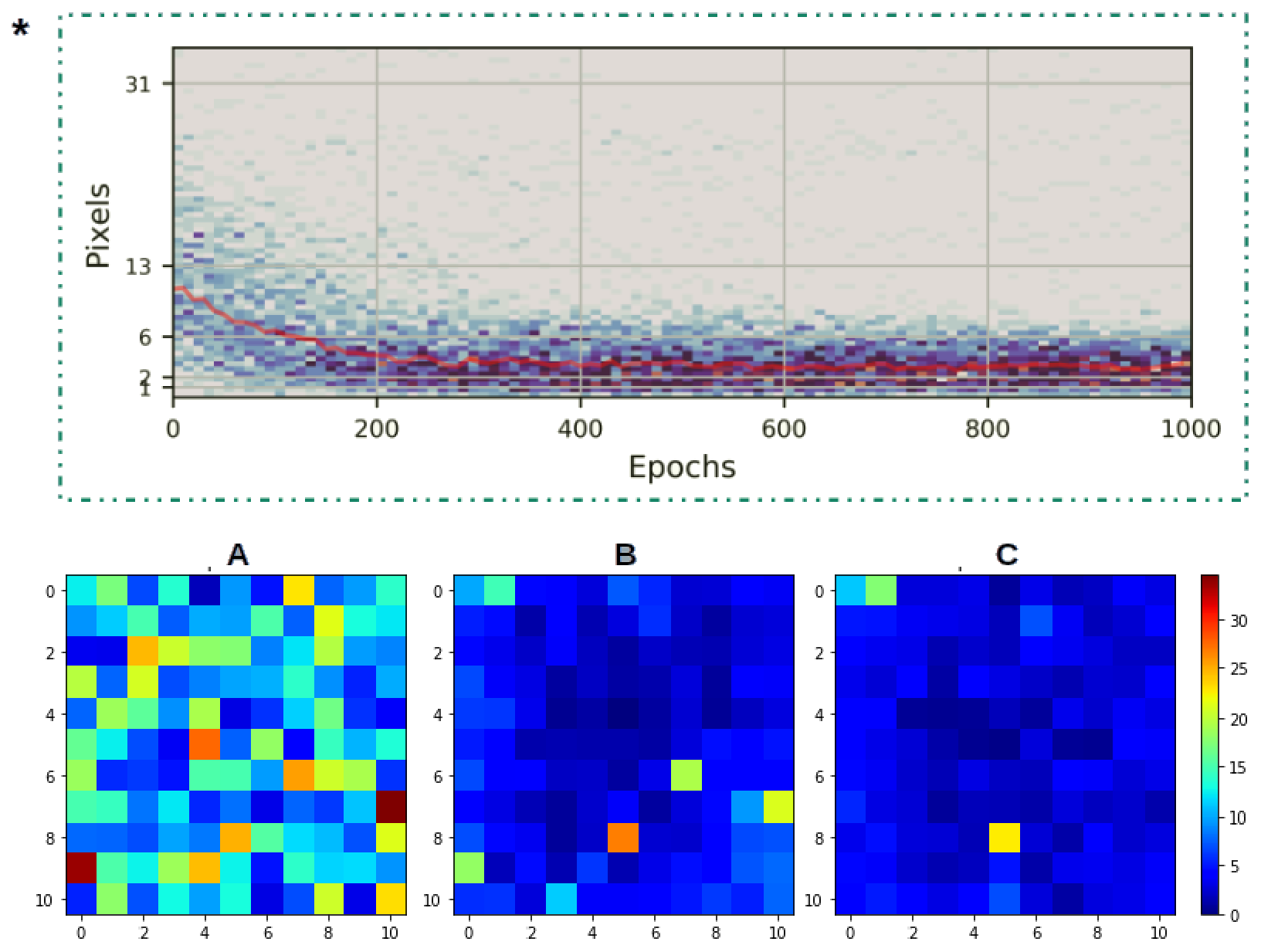}
	\caption{The uppermost graph is the zoomed version of the central panel in Fig. \ref{fig:combinederror} (denoted by an asterisk). Here, the distribution of the positions errors is shown as epochs increase. It can be seen that the median decay towards 0 as the reconstruction proceeds. Panel A,B and C shows the ground truth error for each of the 11x11 positions.}
	\label{fig:poserrcombined}
\end{figure}

\begin{figure}[hbtp]
	\centering
	\includegraphics[width=5 in]{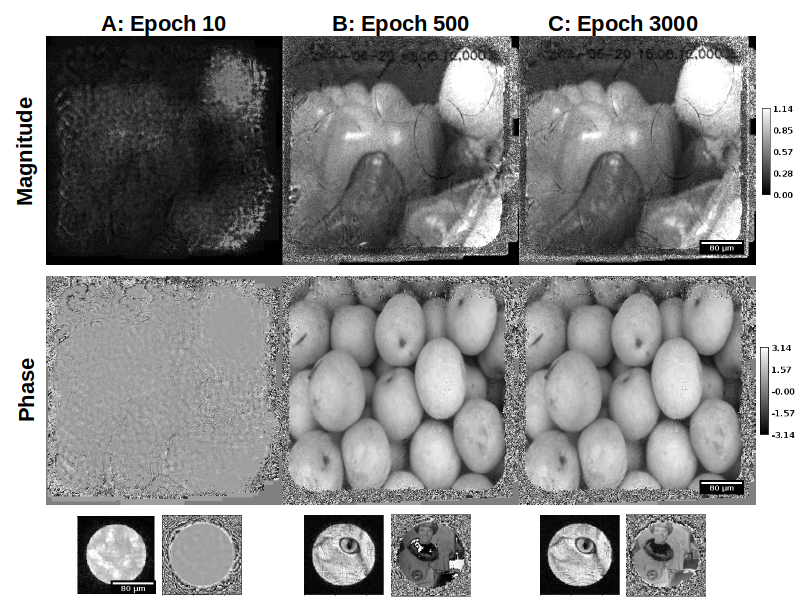}
	\caption{Evolution of the reconstruction for a synthetic dataset, shown at the epoch denoted by A, B and C in Fig. \ref{fig:combinederror}.}
	\label{fig:pepperoni}
\end{figure}

\noindent Fig. \ref{fig:combinederror} and \ref{fig:poserrcombined} show the convergence of the algorithm tested on one example dataset, depicted in Fig. \ref{fig:pepperoni}:  $O(\mathbf{r})$ is made by the "Pepperoni" (magnitude) and "Peaches" (phase) images, while $P(\mathbf{r})$ is composed by the "Chelsea cat" (magnitude) and "Astronaut" (phase) pictures. It can be seen that in around 500 epochs both the positions and the distance have recovered more than 90\% of initial error, producing the object and illumination estimate which can be seen in Fig. \ref{fig:pepperoni}. Note that the during the reconstruction of the illumination (the cat and astronaut insets in Fig. \ref{fig:pepperoni}), no mask is used. The circular pattern is automatically retrieved by the procedure. The loss value (Fig. \ref{fig:combinederror}, top panel) is used as the guiding metrics to optimise the object, the illumination and the setup parameters. The use of the grid sampler is essential to increase the convergence, which tends to be instead slower for the case of optimisation-like reconstruction algorithm based on the a typical crop operator; the acceleration effect can be traced back to the inherent regularisation action of the interpolator.

\noindent In the uppermost panel of Fig. \ref{fig:poserrcombined}, the distribution of the position error is shown for each epoch, denoting a Rayleight-like distribution, which tends to be narrower as the epochs increase; the median, indeed decay towards 0, indicating a good correction. The desired single-modality form of the distribution is an effect of a refined version of the problem expression: we have to optimise only additive correction factors for the positions, not just the positions. In this manner, an $L_1$ metric can be used to create a regularisation term on these correction factors.

\noindent In the bottom panel of Fig. \ref{fig:combinederror} the convergence of the $z$ is shown: similarly to the other two panels in Fig. \ref{fig:combinederror}, the optimisation is moving fast towards the exact value within the first 1000 iterations of the algorithm, denoting how the convergence of each trained variable is aligned with the others, manifesting an ensemble behaviour.

\begin{figure}[htbp]
	\centering
	\includegraphics[width=3.5 in]{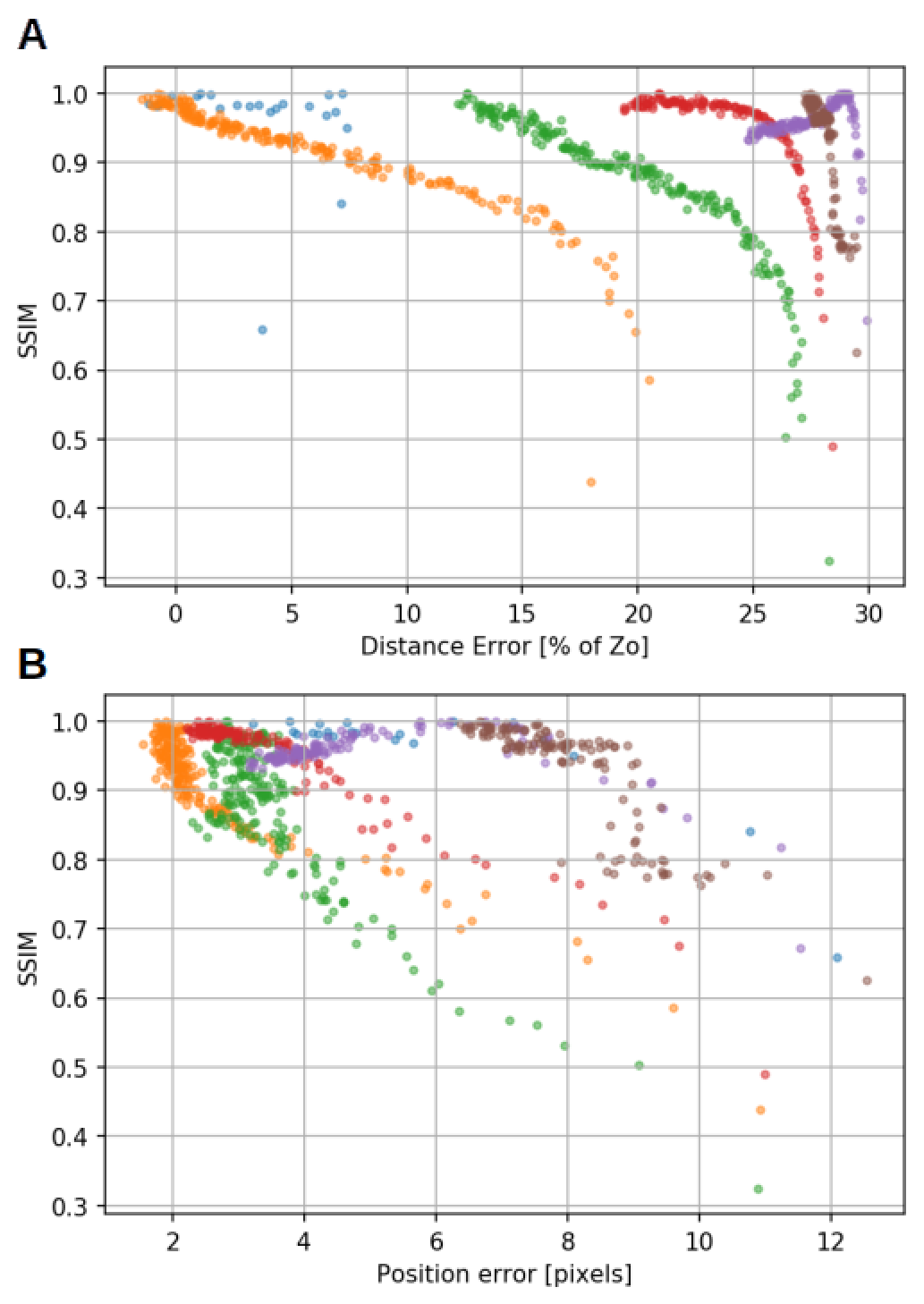}
	\caption{Normalised SSIM as a function of the propagation distance error (panel A) and the median of the scan positions error (panel B), calculated as the reconstruction progresses for many dataset (different colours). Convergence speed in the various phases can be guessed by observing the sparsity of the points for each error value.}
	\label{fig:combinedsyn}
\end{figure}

\noindent To better analyse this aspect, Fig. \ref{fig:combinedsyn} panel A shows how during the reconstruction, the SSIM value increases as the propagation distance (panel A) approaches to the correct value. Each color is relative to a different dataset. The longest trails are the one for which the propagation distance is larger, then the optimisation, initialised with a 30\% error, spans for an extended range of $z$. Conversely, for small propagation distances, the regression of the correct value tends to be faster. Panel B shows a similar graph but for the median of the positions error. 
Convergence speed in the various phases can be guessed by observing the sparsity of the points for each error value.

\section*{Acknowledgments}{We are grateful to Roberto Borghes for its fundamental work on the TwinMic microscope control system. We are also thankful to the System Administrators of the Elettra IT Group, in particular to Iztok Gregori for his work on the HPC solution}

\clearpage
\bibliography{biblio}

\end{document}